\documentclass{aa}

\usepackage{graphicx}
\usepackage{txfonts}
\usepackage{multirow}
\usepackage{amssymb}
\usepackage{units}
\usepackage{float}
\usepackage{natbib}
\bibpunct{(}{)}{;}{a}{}{,}

\usepackage[dvipsnames]{xcolor}
\usepackage{color, colortbl}
\usepackage{xspace}
\usepackage{lscape}
\usepackage{url}
\usepackage{arydshln}

\usepackage[hyperfootnotes=true, linktocpage=true, breaklinks=true, colorlinks=true, linkcolor=magenta, citecolor=blue, urlcolor=blue]{hyperref}
\usepackage[all]{hypcap}

\usepackage{enumitem,booktabs}
\usepackage[referable]{threeparttablex}
\renewlist{tablenotes}{enumerate}{1}

\newcommand{\kms}{\ensuremath{\mathrm{km\ s^{-1}}}\xspace}
\newcommand{\NH}{\ensuremath{N_{\mathrm{H}}}\xspace}

\newcommand{\uhuru}{{\it Uhuru}\xspace}
\newcommand{\xmm}{XMM{\it-Newton}\xspace}
\newcommand{\chandra}{{\it Chandra}\xspace}

\newcommand{\fuse}{{FUSE}\xspace}

\newcommand{\spex}{{\textsc{Spex}}\xspace}

\newcommand{\ergflux}{{\ensuremath{\rm{erg\ cm}^{-2}\ \rm{s}^{-1}}}\xspace}

\newcommand{\degree}{{\ensuremath{^{\circ}}}\xspace}

\newcommand{\cvi}{\ion{C}{vi}\xspace}
\newcommand{\nvi}{\ion{N}{vi}\xspace}
\newcommand{\nvii}{\ion{N}{vii}\xspace}
\newcommand{\neii}{\ion{Ne}{ii}\xspace}
\newcommand{\neiii}{\ion{Ne}{iii}\xspace}
\newcommand{\neix}{\ion{Ne}{ix}\xspace}
\newcommand{\nex}{\ion{Ne}{x}\xspace}
\newcommand{\oii}{\ion{O}{ii}\xspace}
\newcommand{\oiii}{\ion{O}{iii}\xspace}
\newcommand{\ovi}{\ion{O}{vi}\xspace}
\newcommand{\ovii}{\ion{O}{vii}\xspace}
\newcommand{\oviii}{\ion{O}{viii}\xspace}
\newcommand{\fexvii}{\ion{Fe}{xvii}\xspace}

\newcommand{\mgxi}{\ion{Mg}{xi}\xspace}

\newcommand{\foru}{4U~1820-30\xspace}

\defcitealias{Costantini12}{C12}

\definecolor{Gray}{gray}{0.95}
\definecolor{Cyan}{rgb}{0.9,1,1}
\definecolor{Red}{rgb}{0.992, 0.8, 0.69}

\makeatletter
\renewcommand*\aa@pageof{, page \thepage{} of \pageref*{LastPage}}
\setlist[tablenotes]{label=\tnote{\alph*},ref=\alph*,itemsep=\z@,topsep=\z@skip,partopsep=\z@skip,parsep=\z@,itemindent=\z@,labelindent=\tabcolsep,labelsep=.2em,leftmargin=*,align=left,before={\footnotesize}}
\makeatother

\begin{document}

   \title{The hot interstellar medium towards \foru:\\ a Bayesian analysis}

   \titlerunning{The hot interstellar medium towards \foru}
   \authorrunning{D. Rogantini, E. Costantini et al.}

   \author{D. Rogantini
          \inst{1,2},
          E. Costantini\inst{1,2},
          M. Mehdipour\inst{1},
          L. Kuiper\inst{1},
          P. Ranalli\inst{3},
           \and
          L.B.F.M. Waters\inst{1,2}
          }

   \institute{SRON Netherlands Institute for Space Research, Sorbonnelaan 2, 3584 CA Utrecht, the Netherlands\\
              \email{d.rogantini@sron.nl}
             \and 
             Anton Pannekoek Astronomical Institute, University of Amsterdam, P.O. Box 94249, 1090 GE Amsterdam, the Netherlands
             \and
             Combient Mix AB, Box 2150, 40313 Göteborg, Sweden
             }

    \date{Date: \emph{-}}

  \abstract
   {High ionisation lines in the soft X-ray band are generally associated to either interstellar hot gas along the line of sight or to photoionised gas intrinsic to the source. In the low-mass X-ray binary \foru, the nature of these lines is not well understood.}
   {We characterised the ionised gas present along the line of sight towards the source producing the X-ray absorption lines of \mgxi, \neix, \fexvii, \ovii, and \oviii. }
   {We analysed all the observations available for this source in the \xmm and \chandra archives, taken with the RGS, HETG, and LETG spectrometers. The high-resolution grating spectra have been accurately examined through a standard X-ray analysis based on the $C$-statistic and through the Bayesian parameter inference. We tested two physical models which describe a plasma in either collisional ionisation or photoionisation equilibrium. We adopted the Bayesian model comparison to statistically compare the different combinations of the models used for the analysis.}
   {We found that the lines are consistent with hot gas in the interstellar medium rather than being associated to the intrinsic gas of the X-ray binary. Our best-fit model reveals the presence of a collisionally ionised plasma with a temperature of $T=(1.98\pm0.05)\times10^6 $ K. The photoionisation model fails to fit the \fexvii line (which is detected with a significance of $6.5\sigma$) due to the low column density predicted by the model. Moreover, the low inclination of the binary system is likely the reason for the non-detection of ionised gas intrinsic to the source.}
   {}

    \keywords{ 
               X-rays: binaries --
                  X-rays: individuals: 4U 1820-30 --
                    X-rays: ISM --
                    ISM: lines and bands
               }

   \maketitle
%

\section{Introduction}

%
\begin{table*}
\caption{Observation log of \xmm and \chandra data used for our spectral modelling of \foru.}
\label{tab:observations}     
\centering          
\begin{tabular}{ c c c c c c c }     
\hline\hline       
\noalign{\vskip 0.75mm}
Dataset & Obs. ID & Date & Exposure (ks) & Count Rate ($\rm cts\ s^{-1}$) & Instrument & Mode \\ 
\noalign{\vskip 0.75mm}
\hline
\rowcolor{Gray}
\multicolumn{7}{c}{\chandra} \\ 
\hline
\noalign{\vskip 0.75mm}                  
 1                  & \textbf{98}    & $10/03/2000$ & 15.0 & 37.7  & LETG/HRC-S & -   \\ \arrayrulecolor{gray}\cline{2-7}
\noalign{\vskip 0.75mm}     
 \multirow{2}{*}{2} & \textbf{1021}  & $21/07/2001$ & 9.6  & 84.3  & HETG/ACIS  & TE  \\
                    & \textbf{1022}  & $12/09/2001$ & 10.7 & 107.1 & HETG/ACIS  & TE  \\ \cline{2-7}
\noalign{\vskip 0.75mm}    
 \multirow{3}{*}{3} & \textbf{6633}  & $12/08/2006$ & 25.1 & 126.0 & HETG/ACIS  & CC  \\
                    & \textbf{6634}  & $20/10/2006$ & 25.0 & 174.0 & HETG/ACIS  & CC  \\
                    & \textbf{7032}  & $05/11/2006$ & 46.2 & 148.1 & HETG/ACIS  & CC  \\ \cline{2-7}
\noalign{\vskip 0.75mm}
 4                  & \textbf{12444} & $08/03/2011$ & 89.7 & 84.4  & LETG/ACIS  & CC  \\ \arrayrulecolor{black}
\noalign{\vskip 0.75mm}
\hline
\rowcolor{Gray}
\multicolumn{7}{c}{\xmm} \\ 
\hline
\noalign{\vskip 0.75mm}
 5                  & \textbf{008411}0201  & $09/10/2001$ & 39.6 & 74.7 & RGS & Spec. HER+SES  \\ 
 6                  & \textbf{055134}0201  & $02/04/2009$ & 41.8 & 51.3 & RGS & Spec. HER \\ 
\noalign{\vskip 0.75mm}
\hline
\end{tabular}
\end{table*}

In the space among the stars resides interstellar matter, a vast and heterogeneous mixture of atoms, molecules and solid dust grains. As a function of the temperature and density of the diffuse gas, the interstellar medium can be classified in three main phases \citep[][and reference therein]{Ferriere01}: a \emph{cold} phase (with temperature $T_{\rm ISM} < 100$ K and density $n_{\rm H}$ = $10-10^{6}\ \rm cm^{-3}$), a \emph{warm} phase ($T_{\rm ISM} \sim 8000$ K and $n_{\rm H} = 0.2-0.5\ \rm cm^{-3}$, including both warm neutral and warm photoionised gases), and a \emph{hot} phase ($T_{\rm ISM} \sim 10^6$ K and $n_{\rm H} \sim 6.5 \times 10^{-3}\ \rm cm^{-3}$).\\

The existence of the hot interstellar medium (also known as hot coronal gas) in the Milky Way has been discovered through the detection of the soft X-ray background in the 0.1-1 keV energy range \citep{Tanaka77} and through the UV \ovi absorption lines in the spectra of OB stars \citep{Jenkins78}. The hot phase has been proposed to account for much of the missing baryons within the Galactic halo, and as a tracer of energetic feedback from stars and supernovae, which plays an important role in shaping the ecosystem of the Galaxy \citep{Mckee77,Miller13}. \\

In the past several years it has been demonstrated that the hot phase of the ISM can be successfully characterised through X-ray absorption line spectroscopy \citep[e.g.][]{Futamoto04,Yao05,Wang10,Liao13,Gatuzz18}. Through the measurements of absorption lines produced by various ions in the spectra of bright X-ray sources, primarily active galactic nuclei and X-ray binaries, it is possible to set physical constraints on ionisation process, temperature, kinematics, and chemical abundances of the hot gas. It is also possible to study its distribution in both the disc and the halo of the Galaxy. \\

However, the interpretation of these high-ionisation absorption lines can be uncertain. Indeed, these features are not always clearly associated with the interstellar hot gas in a collisional ionisation equilibrium state. In some systems, they can be produced by an absorber intrinsic to the background X-ray source and photoionised by its strong radiation. The presence of multiple blueshifted and/or variable high-ionisation lines indicates a wind outflowing in our line of sight, in particular from the accretion disc of stellar mass black holes \citep[e.g.,][]{Lee02,Parmar02,Ueda04,DiazTrigo06}. On the other hand, in dipping X-ray binaries, strong absorption lines that are local to the source are found not to be blueshifted and these are thought to be associated with the corona of the accretion disc \citep{Sidoli01,DiazTrigo06}. Therefore, for some sources, it might be difficult to discern if the presence of high-ionisation lines in the spectrum is associated with a collisionally ionised gas in interstellar space rather than with a photoionised absorber associated to the X-ray source. One example is the source that we analyse in the present work.\\

\object{4U 1820-30} is a well known bright X-ray source in the Sagitarius constellation, first observed by the \uhuru satellite \citep{Giacconi72}. This source is an accreting neutron star low-mass X-ray binary, residing in the globular cluster NGC 6624 at $(l,b) = (2\degree.8,-7\degree.9)$. The binary consists of an ultracompact system with an orbital period of $\sim 11\ \rm min$ and size $r=1.3\times10^{10}\ \rm cm$ \citep{Stella87}. The companion has been identified with a He-white dwarf. The distance of \foru has been determined to be $7.6\pm0.4\ \rm kpc$ \citep{Kuulkers03}, thus it is very close to the Galactic centre. Consequently, our line of sight samples the entire inner Galactic disk radially to a height of $\sim 1\ \rm{kpc}$ off the Galactic plane. \\

The soft X-ray band of \foru was demonstrated to be a useful tool to study simultaneously the different phases of the interstellar medium. In previous works, the cold phase has been studied through the oxygen, neon, and iron photoelectric edges detected in the high-resolution X-ray spectra \citep{Juett04,Juett06,Miller09}. An accurate analysis of the oxygen K-edge and iron L-edges has been performed by \cite[][hereafter \citetalias{Costantini12}]{Costantini12} where they modelled the absorption by both cold gas and interstellar dust. The detection of low-ionisation lines\footnote{For the wavelength of the lines we refer to \cite{Juett06} and the \spex atomic database \cite{Kaastra18}.}, such as \neii ($14.608$ {\AA}), \neiii ($14.508$ {\AA}), \oii ($23.351$ {\AA}), \oiii ($23.028$ {\AA}) indicates the presence of a low-ionised gas towards \foru with a temperature of $\sim 5 \times 10^{4}\ \rm{K}$ \citep[][\citetalias{Costantini12}]{Cackett08}. \\

The presence of hot gas towards the source has been previously studied. \cite{Futamoto04} clearly detected the \ovii ($21.602$ {\AA}), \oviii ($18.967$ {\AA}) and \neix ($13.447$ {\AA}) absorption lines in the spectrum of \foru. A Gaussian fit to those lines provided estimates for the column densities through the curve of growth analysis. An important shortcoming of such method is that in case of saturated or non-resolved lines, various possible line broadening effects influence the equivalent width of the line and hence the derived column density.  
To tackle this degeneracy, \cite{Yao05} constructed an absorption line model which includes ionisation equilibrium conditions. This model allows studying directly the physical conditions of the absorbing gas, such as temperature and ionic abundances.\\
Both \cite{Futamoto04} and \cite{Yao05} associated these high-ionisation lines to interstellar hot gas under the assumption of a unity filling factor and that all the absorption arises from the same gas. Moreover, the absolute velocity has been found to be consistent with zero, ruling out the possibility of an outflowing gas. \\ 
Subsequently, \cite{Cackett08} modelled the high-ionisation lines in the spectra of \foru using the photoionisation code XSTAR \citep{Bautista01} and they found that photo-ionisation can reproduce the observed lines well suggesting a photoionised absorber, possibly an atmosphere intrinsic to the accretion disc as a plausible explanation. \\
Finally, \cite{Luo14} focused on the \neix and \fexvii absorption lines using a narrow band ($10$ to $15.5$ \AA) of \chandra observations. They found a weak correlation between the equivalent width of the lines and the luminosity of the source. This would suggest that a significant fraction of these X-ray absorbers may originate in the photoionised gas intrinsic to the X-ray binaries. \\

\indent
In the present work, we aim at investigating the origin of these high-ionisation absorption lines. In order to advance previous studies, we apply state-of-art plasma models of the \spex code, with latest accurate atomic databases, to all the \xmm and \chandra spectra of \foru. We test a photoionisation model, a collisionally ionized model and combinations of them. We then analyse the X-ray spectra and compare the models through two different statistical approaches: the standard $C$-statistic \citep{Cash79} and the Bayesian analysis \citep{vanDyk01,Gelman13}. The latter method carefully explores the whole parameter space inferring the most probable estimate of the model parameters.\\
The organisation of the paper is as follows: in Section \ref{sec:obs_datareduction}, we describe the \xmm and \chandra observations and our data reduction procedure. The broadband model of \foru is presented in Section \ref{sec:continuum}. The ionised gas features are analysed and described in Section \ref{sec:hot_gas}, where we also compare the two different models. In Section \ref{sec:discussion} we discuss the nature of the high-ionisation absorption lines seen in the spectra of the accreting source. The summary of the main results of this work is given in Section \ref{sec:summary}. All errors are measured at the 68\% ($1\sigma$) confidence unless marked otherwise.

\section{Observations and data reduction}
\label{sec:obs_datareduction}

4U 1820-30 has been observed by the grating spectrometers aboard on both \chandra and \xmm. \chandra carries two high spectral resolution instruments, namely the High Energy Transmission Grating \citep[HETG,][]{Canizares05} and the Low Energy Transmission Grating \citep[LETG,][]{Brinkman00}. The LETG can operate with either the Advanced CCD Imaging Spectrometer (ACIS) or the High-Resolution Camera (HRC), whereas the HETG can be coupled only with the ACIS. The HETG consists of two grating assemblies, the high-energy grating (HEG) and the medium-energy grating (MEG). The energy resolution of the HEG, MEG and LETG are 0.012, 0.023, 0.05 {\AA} (FWHM), respectively. On the other hand, \xmm carries behind the multi-mirror assemblies two Reflection Grating Spectrometers \citep[RGS,][]{denHerder01}. The energy resolutions of the two instruments, RGS1 and RGS2, are about $0.07$ \AA.\\

In the band of interest of high-ionisation lines, RGS, LETG and HETG have comparable resolving power but different instrumental characteristics. RGS and LETG present higher effective areas and a broader coverage in the softer bands (>10 \AA), which are an optimal combination for detecting in one single observation the absorption lines of highly ionised atoms of multiple elements, down to carbon and nitrogen. HETG best observes higher energy lines such as \neix and \nex. We model both HEG and MEG data. Our joint modelling of \xmm and \chandra data enables us to better constrain all the spectral absorption lines, and therefore to better understand the nature and origin of hot gas towards \foru. The synergy of these different grating instruments improves therefore quantitatively and qualitatively the analysis of the X-ray spectrum of \foru.\\

The characteristics of the observation used in this work are displayed in Table~\ref{tab:observations}. The \xmm data reduction is performed using the Science Analysis System\footnote{\url{https://www.cosmos.esa.int/web/xmm-newton/sas}} (SAS, version 18.0.0). The earlier observation, obsID 008411, shows a flaring background, which is filtered out. This results in a cut of $\sim3$ ks on the total exposure time. The two RGS observations are taken in two different \textit{spectroscopy mode} flavours: high-event-rate with single-event-reconstruction (HER+SES, for obsID 008411) and high-event-rate (HER, for obsID 055134). \\
We obtain the \chandra observations from the Transmission Grating Catalogue\footnote{\url{http://tgcat.mit.edu/}} \citep[TGCat,][]{Huenemoerder11}. We combine the positive and negative first-order dispersion of each observation using the Chandra's data analysis system \citep[CIAO version 4.11,][]{Fruscione06}. The two observations taken in 2001 in timed exposure (TE) mode, display similar fluxes and continuum parameters, therefore we stack the data, for the HEG and MEG arms using the CIAO tool \texttt{combine\_grating\_spectra}. Similarly, we combine the three observations taken in 2006 in continuous clocking (CC) mode. We fit with the same continuum model the HEG and MEG spectra of the same stacked data correcting them for the slightly different ($\sim$ 5\%) instrumental normalization. Therefore, after the combination of the grating spectra, we obtain six different datasets as listed in the first column of Table \ref{tab:observations}.\\

During all the observations, the source is observed in high flux state, on average $ F_{\rm 2-10\ keV} \sim 10^{-8}\ \ergflux$. Because of this high flux, the observations are affected by pileup. This effect is particularly present in the \chandra observations taken in TE mode. We overcome this issue ignoring the region of the spectra ($\lambda$ between 4 and 19 \AA) most affected by pileup. At longer wavelengths ($\lambda > 20 $ \AA), the combined effect of high absorption and low effective area reduces significantly the observed flux. Thus, this region is unaffected by pileup and can be safely used. However, the narrow absorption features are in general minimally influenced by pileup.

   \begin{figure}
   \centering
     \includegraphics[width=\columnwidth]{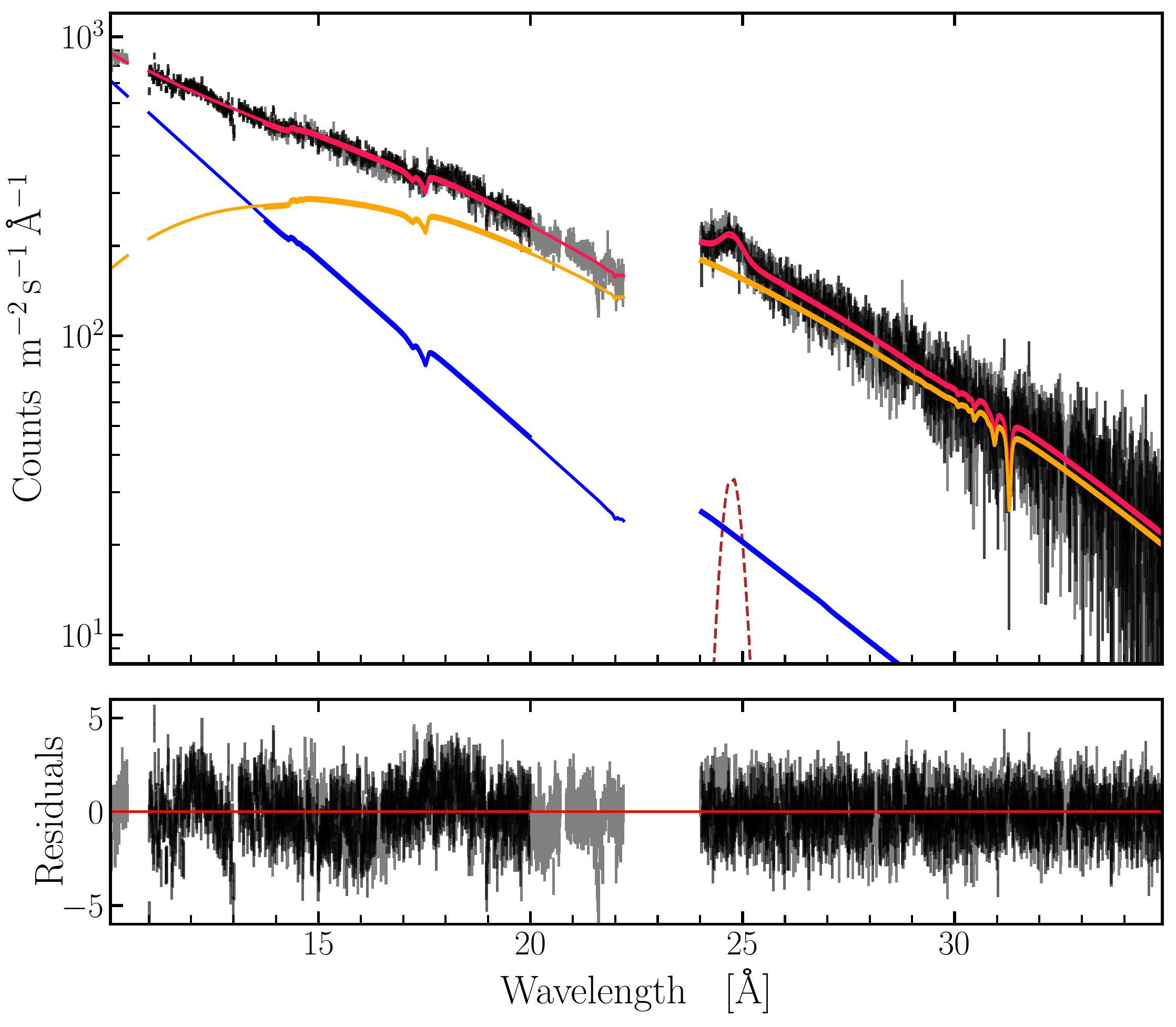}
      \caption{Broadband best-fit to the X-ray spectrum of 4U 1820-30 with warm and cold absorption described in Section \ref{sec:continuum}. We display for clarity only RGS1 and RGS2 (in gray and black, respectively) of obsID 008411. We overlap the best model of the broadband spectrum (in red), which is made up of a Comptonisation model (in blue) and a blackbody (in orange). We also show the gaussian (in brown) added to the model in order to fit the possible instrumental bump in excess centred at $24.7$ \AA. We divide the data for the instrumental effective area in order to clean the spectrum from the numerous instrumental features. In the bottom panel we show the residuals defined as $\rm (observation-model)/error$.}
         \label{fig:continuum}
   \end{figure}

\section{Broadband spectrum}
\label{sec:continuum}

%
\begin{table*}
\caption{Best-fit parameters of our model for the broadband spectrum of 4U 1820-30}
\label{tab:broadband}
\small    
\centering
\begin{threeparttable}        
\begin{tabular}{ c c c c c c c c c }     
\hline\hline       
\noalign{\vskip 0.75mm}
\multirow{3}{*}{data} & {\tt bb} &  \multicolumn{2}{c}{\tt comt\tnotex{tnote:tau} } & {\tt hot} (cold) \tnotex{tnote:cold} & \multicolumn{2}{c}{{\tt hot} (warm)} & \multirow{1}{*}{$F_{0.5-2\ \rm keV}$} & \multirow{3}{*}{$C$stat/dof\tnotex{tnote:tot_c}} \\ \cline{3-4}\cline{6-7}
\noalign{\vskip 0.75mm}
        & $k_{\rm B}T$ & $k_{\rm B}T_{0}$ & $k_{\rm B}T_{1}$ & \NH  & \NH & $k_{\rm B}T$ & {\scriptsize$[10^{-9}]$}   \\
\noalign{\vskip 0.75mm}
        & \tiny keV & \tiny keV & \tiny keV & \tiny $\rm 10^{21}\ cm^{-2}$ & \tiny $\rm 10^{20}\ cm^{-2}$ & \tiny $\rm 10^{-3}\ keV$ & \tiny \ergflux &   \\

\noalign{\vskip 0.75mm}
\hline
\rowcolor{Gray}
\multicolumn{9}{c}{RGS/\xmm} \\
\hline
\noalign{\vskip 0.75mm}              
008411   & $ 0.164\pm0.003 $ & $ 0.65\pm0.08 $ & $24\pm11    $ & $ 1.51\pm0.03$ & $ 0.9\pm0.2$ & $ 2.6\pm0.3$ & $ 2.2\pm0.2$ & $\bf5642/3700$ \\
055134   & $ 0.139\pm0.004 $ & $ 0.32\pm0.02 $ & $29\pm9     $ & $ 1.56\pm0.03$ & $ 1.2\pm0.3$ & $ 2.8\pm0.2$ & $ 1.5\pm0.3$ & $\bf5541/3921$ \\
\noalign{\vskip 0.75mm}
\hline
\rowcolor{Gray}
\multicolumn{9}{c}{LETG/\chandra} \\
\hline
\noalign{\vskip 0.75mm} 
98    & $ 0.16\pm0.01   $ & $ 0.40\pm0.08 $ & $7\pm5      $ & $ 0.08\pm0.05$ & $ <0.45    $ & $ 2\pm1    $ & $ 1.8\pm0.8$ & $\bf2070/1842$ \\
12444 & $ 0.22\pm0.01   $ & $ 1.9\pm0.7   $ & $20\pm9     $ & $ 1.4\pm0.1$   & $ 2.0\pm0.9$ & $ 4.6\pm0.4$ & $ 1.5\pm0.7$ & $\bf1141/870$ \\
\noalign{\vskip 0.75mm}
\hline
\rowcolor{Gray}
\multicolumn{9}{c}{HETG/\chandra} \\
\hline
\noalign{\vskip 0.75mm}
2001 \tnotex{tnote:2001}  & $ 0.14\pm0.04   $ & $ 0.20\pm0.05 $ & $1.5\pm0.2  $ & $ 1.1\pm0.2$   & $ 1.5\pm0.9$ & $ 4\pm1    $ &  $ 1.7\pm0.2$ & $\bf7423/6628$ \\
2006 \tnotex{tnote:2006}  & $ 0.07\pm0.01   $ & $ 0.16\pm0.02 $ & $9\pm5      $ & $ 2.1\pm0.1$   & $ 1.2\pm0.9$ & $ 3\pm1    $ &  $ 2.3\pm0.3$ & $\bf6304/4988$ \\
\noalign{\vskip 0.75mm}
\hline  
\noalign{\vskip 0.75mm}
Mean    &    $0.15\pm0.01$  & $0.6\pm 0.2 $ &   $15\pm7$      &   $1.53\pm0.08$  &   $1.4\pm0.6$  & $ 3.4\pm0.6 $ &  $10\pm2$ &    \\

\noalign{\vskip 0.75mm}
\hline
\end{tabular}
\begin{tablenotes}
      \item\label{tnote:tau}{\tiny We fix the optical depth $\tau$ to its default value of 3.}
      \item\label{tnote:cold}{\tiny We fix the temperature at the lower limit $0.5$ eV. In the cold phase we also consider the contribution of the dust.}
      \item\label{tnote:tot_c}{\tiny The total $C$-stat/dof of the fit is $28123/21949$}
      \item\label{tnote:2001}{\tiny Combination of obsId (1021, 1022) taken in 2001.}
      \item\label{tnote:2006}{\tiny Combination of obsId (6633, 6634 and 7032) taken in 2006.}

\end{tablenotes}
\end{threeparttable}
\end{table*}

We fit the HETG, LETG and RGS data simultaneously using the \spex X-ray spectral fitting package\footnote{\url{http://doi.org/10.5281/zenodo.2419563}} \citep[Version 3.05.00,][]{Kaastra18} with the $C$-statistic. Considering all the datasets, the observations cover the soft X-ray energy band between 4 and 35 {\AA} ($0.35 - 3.10$ keV). Given that the source was observed at different epochs, the continuum shape may differ significantly. To take into account continuum variability, we assign each dataset to a specific \texttt{sector} in \spex allowing the continuum parameters to vary freely for each sector. The broadband model with the cold and warm absorptions (described below) is fitted to the data adopting the \emph{C}-statistic test (see Figure \ref{fig:continuum}).\\

The source shows a Comptonised continuum together with a soft black-body component \citep[e.g.,][\citetalias{Costantini12}]{Sidoli01b}, which we reproduce in our modelling using the \texttt{comt} and \texttt{bb} components in \spex \citep[][respectively]{Titarchuk94,Kirchhoff60}. We first characterise the absorption by the neutral ISM gas. This is done by applying a collisionally ionised plasma model \citep[model \texttt{hot} in \spex,][]{dePlaa04,Steenbrugge05} with a temperature frozen to $k_{B}T= 0.5\ \rm eV$ in order to mimic the neutral gas. Continuum absorption by this neutral Galactic absorber takes into account most of the low-energy curvature of the spectrum especially for $\lambda \gtrsim 6 $ {\AA} ($E\lesssim 2$ keV). Due to the complexity of the oxygen K-edge region and to the uncertainties of its modelling we exclude from the fit the region extending between $22.2-24$ {\AA} ($\sim 516-558$ eV). Modelling this complex edge is beyond the scope of this paper.\\
Three other photoabsorption edges are present in the spectrum of the source\footnote{For the wavelength of the photoabsorption edges we refer to \cite{Gorczyca00} and \cite{Kortright00}.}: the neon K-edge at $\lambda = 14.3$ {\AA}, the iron L$_{2,3}$-edges at $\lambda_{2,3} = 17.2,\ 17.5 $ {\AA} and the nitrogen K-edge at $\lambda = 31.1$ {\AA}. While neon, nitrogen are expected to be mostly in gaseous form (especially neon due to its inert nature), iron is highly depleted from the gaseous form \citepalias{Costantini12}. To minimise the residual in the Fe L-edge band and to characterise the dust contribution, we add to our model the \texttt{AMOL} component \citep{Pinto10}. In particular, we adopt the Fe L-edges of the metallic iron. Coupling the abundance of the dust among all the datasets, we find that $(92\pm4)\%$ of the iron is locked into interstellar dust grains.\\
The detection of the \neii, \neiii lines at 14.608 {\AA}, 14.508 {\AA} (848.74 eV and 854.59 eV), respectively, traces the presence of low-ionised gas along this line of sight. Therefore, to describe these low-ionisation lines we add a second collisionally ionised plasma model to our broadband model. \\
The best fit parameters of the broadband continuum including the cold and warm absorption are shown in Table \ref{tab:broadband}. The different fit of LETG obsID 98 might be caused by residuals at the metal edges, in particular the O K-edge, due to uncertainties in the \chandra response matrix \citep{Nicastro05}. Thus, the estimates of the hydrogen column densities are altered by these residuals and we do not consider them in the average computations. However, the high-ionisation absorption lines are not affected and we can reliably consider them in our analysis.\\
We also test the effect of dust scattering using the \texttt{ismdust} model developed by \cite{Corrales16} and implemented into Xspec \citep{Arnaud96}. Adding this component to the continuum model, we obtain about 10\% lower hydrogen column density.\\

\section{High-ionisation lines}
\label{sec:hot_gas}

The spectra of \foru display several lines from highly ionized elements: \mgxi, \neix, \fexvii, \ovii and \oviii. In Table \ref{tab:high_lines}, we list all the lines with their detection significance. The effective area of RGS and LETG HRC should allow us to detect the absorption lines of \cvi, \nvi and \nvii. However, since in this region the cold-gas absorption of the continuum is severe, the signal-to-noise ratio is too low for a meaningful analysis. \\
\begin{table}
\caption{High-ionisation lines detected in the \xmm and \chandra spectrum.}
\label{tab:high_lines}     
\centering          
\begin{tabular}{ c c c c }     
\hline\hline       
\noalign{\vskip 0.75mm}
Line & $\lambda$ [\AA] & $E$ [keV] & Significance \\
\noalign{\vskip 0.75mm}
\hline
\noalign{\vskip 0.75mm}                  
\mgxi              & $9.1688$ & $1.3522$ & $ 6.5\sigma$  \\
\neix              & $13.447$ & $0.9220$ & $ 3.9\sigma$  \\ 
\fexvii            & $15.012$ & $0.8259$ & $ 10\sigma$   \\ 
\ovii He$\alpha$   & $21.590$ & $0.5743$ & $ 4.6\sigma$  \\ 
\ovii He$\beta$    & $18.626$ & $0.6657$ & $ 3.1\sigma$  \\ 
\oviii Ly$\alpha$  & $18.967$ & $0.6537$ & $ 8\sigma$    \\ 
\oviii Ly$\beta$   & $16.005$ & $0.7747$ & $ 11\sigma$   \\
\cvi               & $33.734$ & $0.3675$ & $ <1\sigma$   \\
\nvi               & $28.788$ & $0.4307$ & $ <1\sigma$   \\
\nvii              & $24.779$ & $0.5004$ & $ 1.8\sigma$  \\ 
\noalign{\vskip 0.75mm}
\hline
\end{tabular}
\end{table}
\noindent
These lines are produced by ionised plasma present along the line of sight of the source. To reveal the nature of this ionised gas we study the associated absorption features through two different physical models defined by us as \textit{ism} and \textit{photo} models:\\
\indent
{\textit{ism model}} -  To model the absorption by the hot coronal gas in the interstellar medium we use the collisionally ionised plasma model \texttt{hot} with a temperature above $10^{5.5}$ K ($E\gtrsim 0.05 $ keV).\\
\indent
{\textit{photo model}} -  To investigate the occurrence of photoionised gas intrinsic to the binary, we adopt the \texttt{xabs} model in \spex \citep{Steenbrugge03}. The \texttt{xabs} model calculates the transmission through a slab of photoionised gas where all ionic column densities are linked in a physically consistent fashion through a photoionisation model. To compute the ionisation balance for the \texttt{xabs} model, we use the photoionisation model of \spex, called \texttt{PION} \citep{Mehdipour16}. This was done for each dataset independently. For this calculation we adopt the proto-solar abundances of \cite{Lodders10}. We retrieve the spectral energy distribution (SED) for each dataset normalising the SED of \citetalias{Costantini12} to the soft X-ray band of the observations. In this energy band, the spectral shape remains consistent and only the normalisation varies. Since our observations do not cover the hard X-ray band, we assume that the shape of SED does not vary and remains consistent to the SED of \citetalias{Costantini12}. The \texttt{PION} calculations yield temperature and ionic column densities as a function of the ionisation parameter $\xi$ \citep{Tarter69}, which are used for fitting with the \texttt{xabs} model. The ionisation state of the absorber is measured through the ionisation parameter, which is defined as:
\[
\xi = \frac{L_{\rm ion}}{n_{\rm{H}} r^2}\ ,
\]
where $L_{\rm ion}$ is the source ionisation luminosity between 1 and 1000 Ryd, $n_{\rm{H}}$ the hydrogen density of the absorber and $r$ its distance from the ionising source. \\

We fit the two models using two different approaches: first the $C$-statistic, widely used in X-ray data analysis, and then the Bayesian parameter inference \citep{Gregory05,Gelman13}. The $C$-statistic can have some difficulties to identify multiple, separate, adequate solutions (i.e. local probability maxima) in the parameter space. On the contrary, the Bayesian approach explores the entire parameter space identifying the sub-volumes which constitute the bulk of the probability. Moreover, it performs optimisation and error estimation simultaneously, but requires large computation time depending on the number of free parameters. For each model, we evaluate the equivalent hydrogen column density (\NH), the flow- ($v$) and turbulence-velocities ($\sigma_{\rm v}$ or velocity dispersion) of the absorber, together with the temperature ($k_{\rm B}T$) or the ionisation parameter ($\xi$), depending on the model.\\

   \begin{figure*}
   \centering
     \includegraphics[width=.8\textwidth]{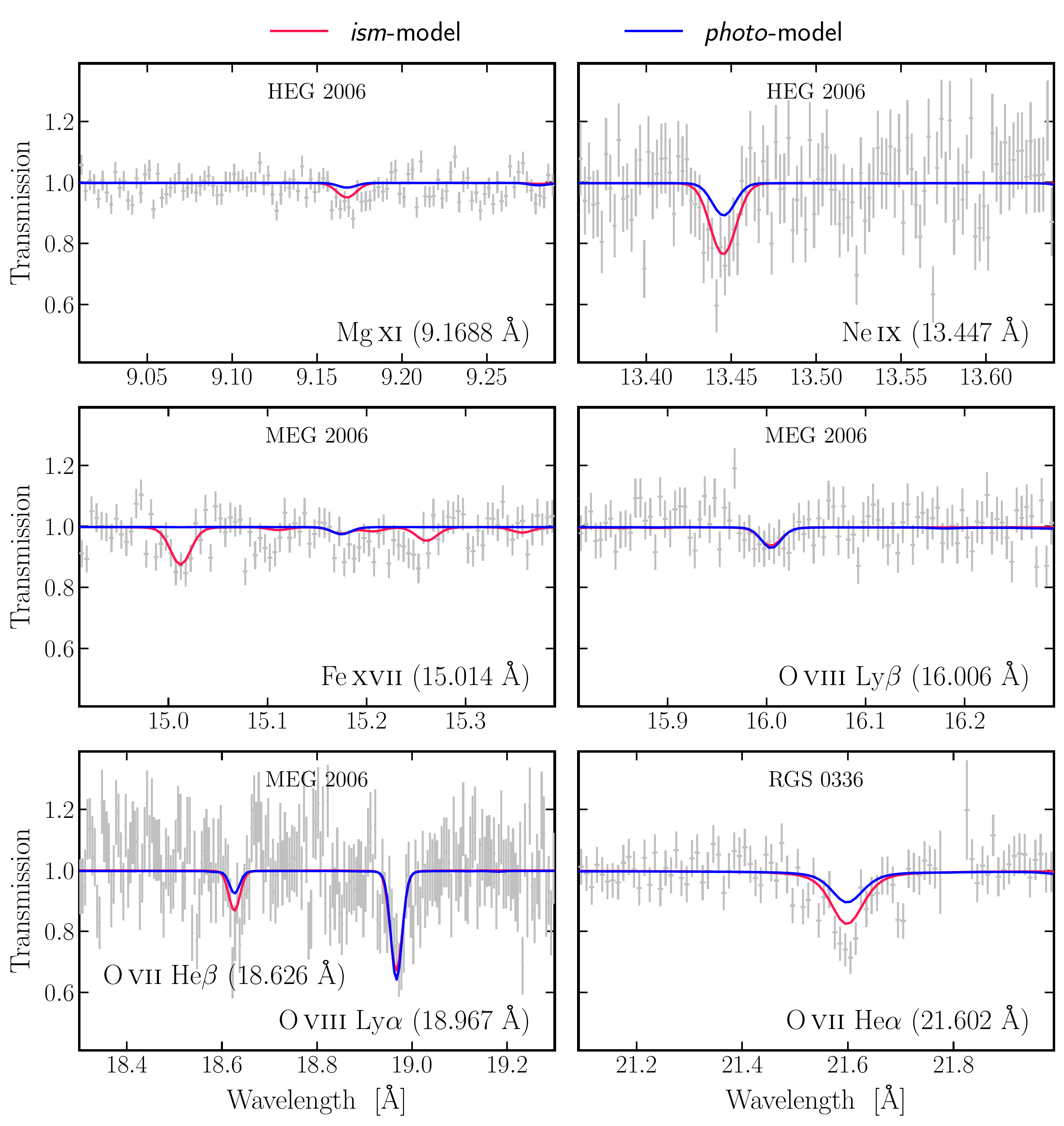}
      \caption{High ionisation lines detected in the spectra of \foru. Here, the spectrum is shown in unit of transmittance: the observed counts are divided by the underlying continuum together with the cold and warm absorption. For clarity, we do not display all the datasets. We superimpose the \emph{ism} and \emph{photo} models (in red and blue, respectively) obtained with the Bayesian parameter inference.}
         \label{fig:all_lines}
   \end{figure*}

\subsection{\textit{C}-statistic analysis}
\label{sec:cstat}

\begin{figure}
   \centering
     \includegraphics[width=\columnwidth]{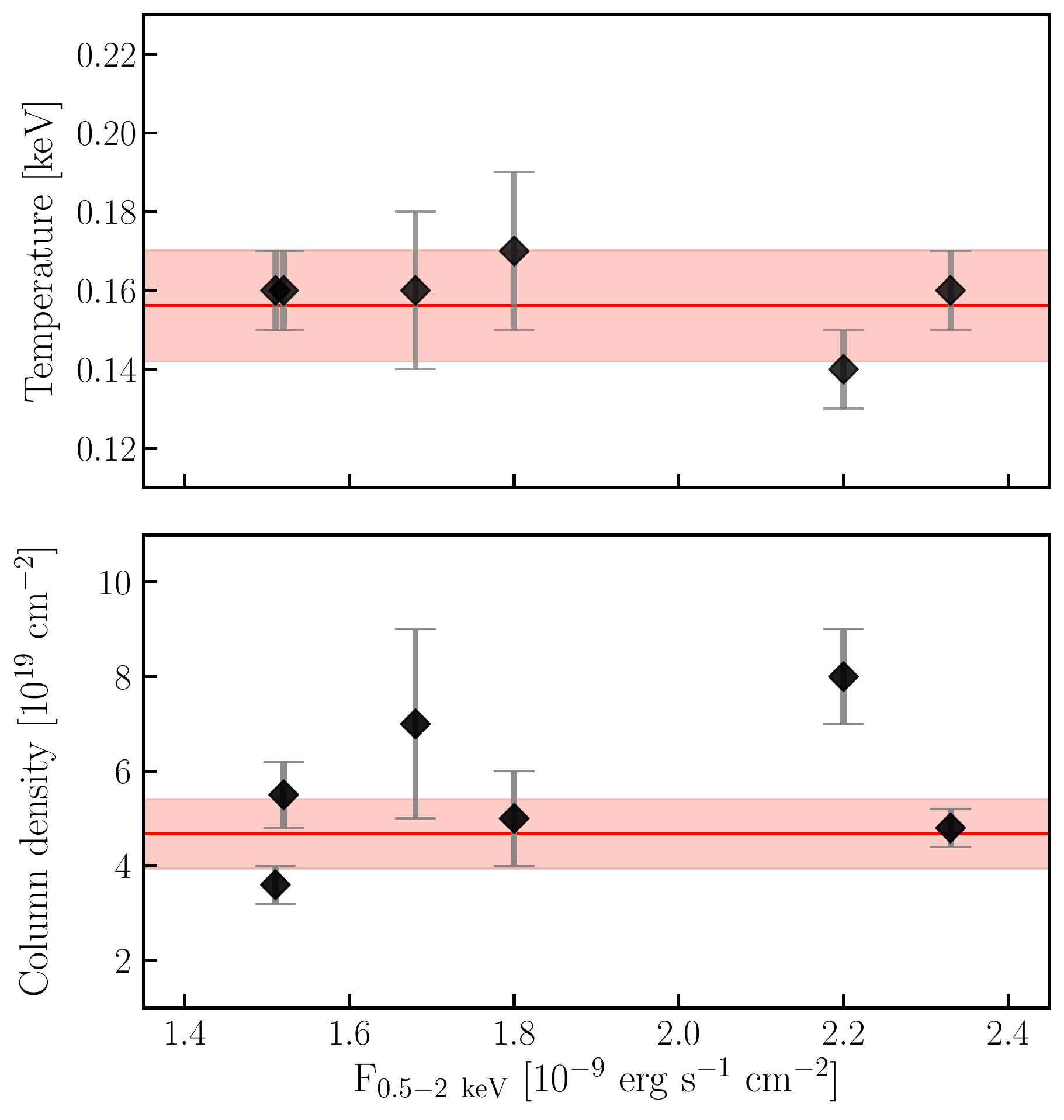}
      \caption{Temperature and the hydrogen column density of the $ism$-model versus the $0.5-2$ keV unabsorbed flux for each dataset using the $C$-statistic fitting model. The horizontal red line represents the inverse-variance weighted average and the coloured band indicates the $3\sigma$ confidence band. Only RGS 008411 (the observation with a flux $\sim2.2\times 10^{-9}\ \ergflux$), shows a deviating column density value.}
         \label{fig:variance}
   \end{figure}

For each dataset, we apply separately the \textit{ism} and \textit{photo} models to the broadband continuum model. We constrain all the free parameters of the broadband fit to within $1\sigma$ uncertainties. We leave only the normalisations of the blackbody and Comptonisation components free to vary (see Section \ref{tab:broadband}). Then, we fit the \texttt{hot} and \texttt{xabs} parameters separately for each dataset and we report their best values in Table \ref{tab:xabs_hot} with the relative uncertainties. \\
For each epoch, a collisionally ionised gas with an average temperature of $0.16\pm0.01$ keV ($ 1.9\pm0.1 \times 10^6$ K) better represents (with a total $\Delta C \rm{stat} = 312$) the high-ionisation absorption lines than a photoionised gas with an average ionisation parameter $\log \xi = 1.72\pm0.05$. We do not observe any significant flow velocity ($v$) associated to the absorber. The estimates of $v$ are, indeed, consistent within the uncertainties with a static gas. For RGS and LETG observations, we fix the turbulent velocities to their default value of $100$ km/s to tackle the degeneration between the fit parameters (in particular between $\sigma_{\rm v}$ and $N_{\rm H}$). Only with HETG, we are able to resolve the turbulent and flow velocities of the absorber. Moreover, the tabulated parameter, such as temperature and \NH, do not show a significant variability among the different epochs, as shown in Figure \ref{fig:variance}.\\
Furthermore, we fit the high-ionisation lines with both the \textit{ism} and \textit{photo} models to test the possible coexistence of photoionised and collisionally ionised gases along the line of sight. In this scenario the $ism$-model dominates the fit and the contribution of the photoionised gas is negligible for each epoch analysed. Statistically, the fit does not improve significantly to support the coexistence of the two absorbers. For the $ism+photo$ model, we find a $C$-stat/dof = 27032/21906 with a tiny improvement in the $C$-stat ($\Delta C = 8 $) with respect to the fit with the $ism$ model alone. 

%
\begin{table*}[t]
\caption{Comparison of the photo-ionised and collisional-ionised best-fit models parameters obtained through the $C$-statistic analysis.}
\label{tab:xabs_hot}
\tiny     
\centering          
\begin{tabular}{ c c c c c c @{\hspace{0mm}}c@{\hspace{12mm}} c c c c c }     
\hline\hline       
\noalign{\vskip 0.75mm}
\multirow{3}{*}{dataset} & \multicolumn{4}{c}{{\it ism} model (\texttt{hot})} & \multirow{3}{*}{$C$stat/dof} & & \multicolumn{4}{c}{{\it photo} model (\texttt{xabs})} & \multirow{3}{*}{$C$stat/dof} \\\cline{2-5}\cline{8-11}
\noalign{\vskip 0.5mm}
       & \NH & $k_{\rm B}T$ & $\sigma_{\rm v}$ & $v$  & &    & \NH & $\log \xi$ & $\sigma_{\rm v}$ & $v$ &    \\
       & $10^{19}\rm cm^{-2}$ & keV & km/s & km/s &  &   & $10^{19}\rm cm^{-2}$ &  & km/s  & km/s &   \\
\noalign{\vskip 0.75mm}
\hline
\rowcolor{Gray}
\multicolumn{12}{c}{RGS/\xmm} \\
\hline
\noalign{\vskip 0.75mm}                    
008411   & $ 8\pm1$   & $0.14\pm0.01$ & $ 100\ (f) $           & $-60\pm40 $  & $\bf5379/3697$&     & $ 17\pm2$       & $1.86\pm0.01 $ & $100\ (f) $      & $-70^{+30}_{-70} $   & $\bf5429/3697$\\
\noalign{\vskip 0.5mm} 
055134   & $ 5.5\pm0.7$ & $0.16\pm0.01$ & $ 100\ (f)$      & $50^{+40}_{-90} $       & $\bf5370/3918$&   & $ 7\pm 1$ & $1.72\pm0.03 $ & $100\ (f)  $ & $50\pm^{+40}_{-80}$ &  $\bf5393/3918$\\
\noalign{\vskip 0.75mm}
\hline
\rowcolor{Gray}
\multicolumn{12}{c}{LETG/\chandra} \\
\hline
\noalign{\vskip 0.75mm}
98    & $ 5\pm1$ & $0.17\pm0.02$ & $ 100\ (f) $          & $20\pm60 $              & $\bf1994/1839$ &  & $ 5\pm1$ & $1.66\pm0.11 $ & $100\ (f) $  & $70\pm{80}     $  & $\bf2013/1839$ \\
\noalign{\vskip 0.5mm} 
12444 & $ 3.6\pm0.4$ & $0.16\pm0.01$ & $  100\ (f) $     & $-190\pm40 $            & $\bf1004/866$ &  & $ 4.2\pm0.6$ & $1.57\pm0.09 $ & $100\ (f) $ & $-160\pm{50} $     & $\bf1042/866$ \\
\noalign{\vskip 0.75mm}
\hline
\rowcolor{Gray}
\multicolumn{12}{c}{HETG/\chandra} \\
\hline
\noalign{\vskip 0.75mm}
2001  & $ 7\pm2$   & $0.16\pm0.02$ & $ 40\pm30 $ & $-60\pm30 $               & $\bf7328/6624$ &  & $ 10\pm2$   & $1.74\pm0.02 $ & $50\pm20   $ & $-60\pm30       $  & $\bf7362/6624$ \\
\noalign{\vskip 0.5mm} 
2006  & $ 4.8\pm0.4$ & $0.16\pm0.01$ & $ 150\pm20$        & $-10^{+10}_{-20} $         & $\bf5962/4984$ &  & $ 6.5\pm0.6$   & $1.72\pm0.01 $ & $130^{+20}_{-100}  $ & $20^{+10}_{-40}        $  &  $\bf6113/4984$ \\
\noalign{\vskip 0.75mm} 
\hline
\noalign{\vskip 0.75mm}     
Mean    &   $5.7\pm0.9$    &  $0.16\pm0.01$ & $95\pm25$ & $-40^{+40}_{-50}$ &  &  & $8\pm1$   &     $1.72\pm0.05$     & $90^{+20}_{-60}$ & $-25^{+40}_{-60}$ &   \\
\noalign{\vskip 0.75mm}
\hline
\end{tabular}
\tablefoot{We indicate with $(f)$ the frozen parameters of the model. The total $C$-stat/dof for the $ism$-model is $27040/21928$ whereas for the $photo$-model is $27352/21928$.}
\end{table*}

\subsection{Bayesian analysis}
\label{sec:bayesian}

   \begin{figure*}[h]
   \centering
     \includegraphics[width=.75\textwidth]{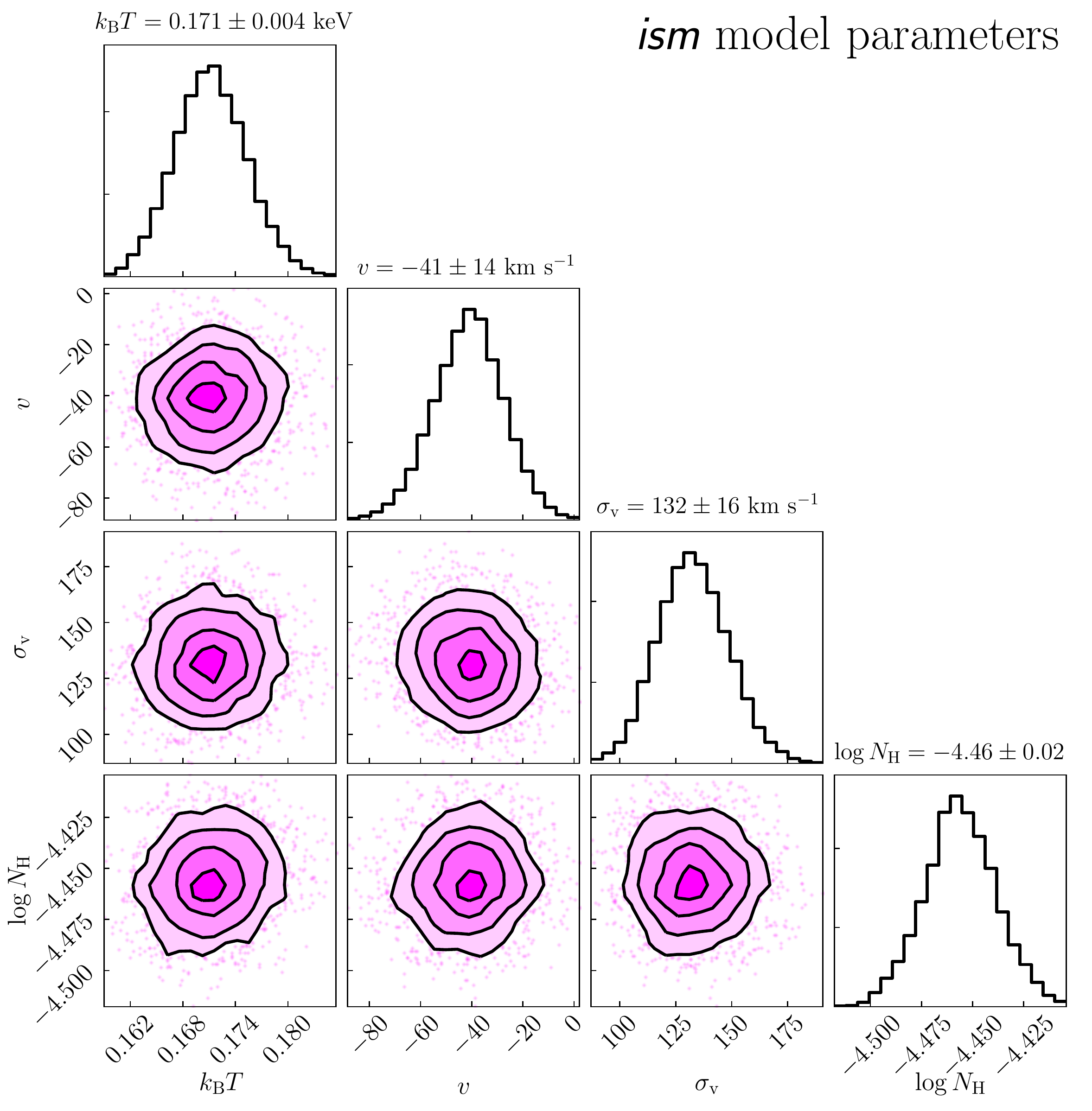}
      \caption{The posterior distribution from MultiNest is plotted with two-dimensional histograms comparing each pair of free parameters \texttt{hot}-model (\NH, $k_{\rm B}T$, $v$ and $\sigma_{v}$). The contours indicate the $1 \sigma$, $2 \sigma$, $3 \sigma$ and $4 \sigma$ confidence intervals in two dimensional space. The hydrogen column density \NH is expressed in unit of $10^{24} cm^{-2}$. }
         \label{fig:hot_dist}
   \end{figure*}

   \begin{figure*}[h]
   \centering
     \includegraphics[width=\textwidth]{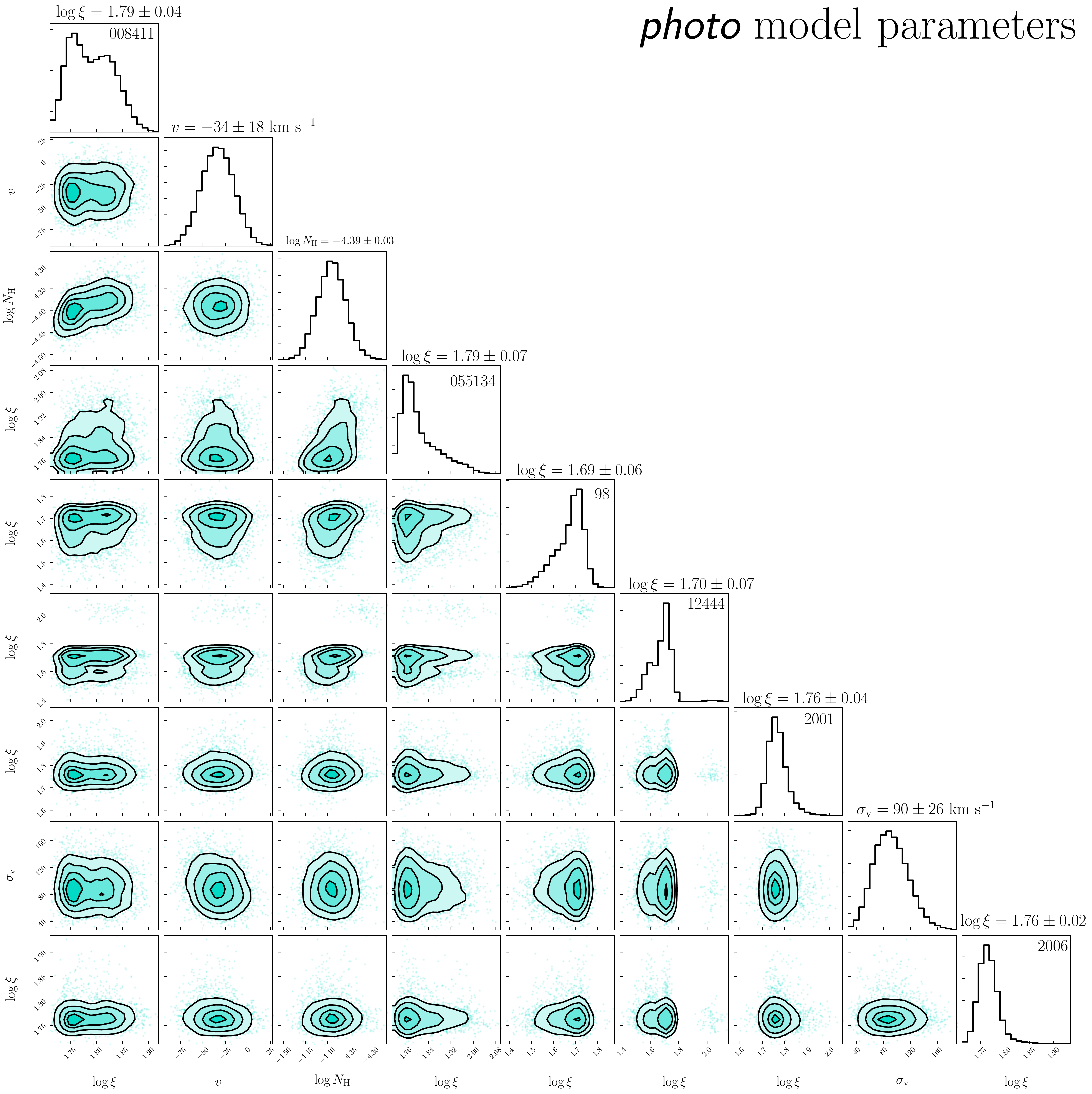}
      \caption{The posterior distribution from MultiNest is plotted with two-dimensional histograms comparing each pair of free parameters of the \texttt{xabs}-model (\NH, $v$, $\sigma_{v}$ and the $\log \xi$ for each observation. The number of observation is indicated in the distribution panel). The contours indicate the $1 \sigma$, $2 \sigma$, $3 \sigma$ and $4 \sigma$ confidence intervals in two dimensional space. The hydrogen column density \NH is expressed in unit of $10^{24} cm^{-2}$.}
         \label{fig:xabs_dist}
   \end{figure*}

For the spectral analysis using the Bayesian approach \citep{Bayes63}, we adopt the MultiNest algorithm \citep[version 3.10,][]{Feroz09,Feroz13}. This method is applicable to low-dimensional problems of X-ray spectral modelling. Its strength is the capability to identify local maxima without difficulty computing points of equal weighting similar to a Markov Chain. It provides values, error estimates and marginal probability distributions for each parameter. \\
To connect the X-ray fitting code \spex with the Bayesian methodology, we create a Python package \textsc{BaySpex}\footnote{\textsc{BaySpex} is publicly available on \url{ZenodoLINK}.}, which is a simplified and adapted version of the \texttt{PyMultiNest} and Bayesian X-ray Analysis (\texttt{BXA}) developed by \cite{Buchner14}. This script is adapted to PYSPEX, the python interface to \spex. The logic behind the script is that MultiNest suggests parameters on a unit hypercube which are translated into model parameters, readable by \spex, using the prior definitions. At this point, \textsc{BaySpex} computes a probability using the \spex likelihood implementation, which is passed back to the MultiNest algorithm.\\

\subsubsection{Parameter inference}
\label{sec:parameter_inf}
Similar to the $C$-statistic approach, we separately multiply the broadband continuum fit with the \emph{ism} and \emph{photo} models. To minimise the number of the free parameters in the fit, we freeze the broadband continuum shape calculated in Section \ref{sec:continuum}. Furthermore, we impose the parameters of the $ism$ model to be the same for all the datasets. Instead, for the $photo$ model, we fit separately the ionisation parameters of the different epochs since each dataset uses a different SED. We couple the velocities and the hydrogen column density to keep the model simple. The Bayesian data analysis is indeed limited by the computation power and time. Moreover, these arrangements are justified by the previous $C$-statistic analysis. The \texttt{hot} and \texttt{xabs} components do not modify the shape of the broadband model and their parameters are constant along all the epochs (see Table \ref{tab:xabs_hot}).\\

\begin{figure*}[h]
   \centering
     \includegraphics[width=\textwidth]{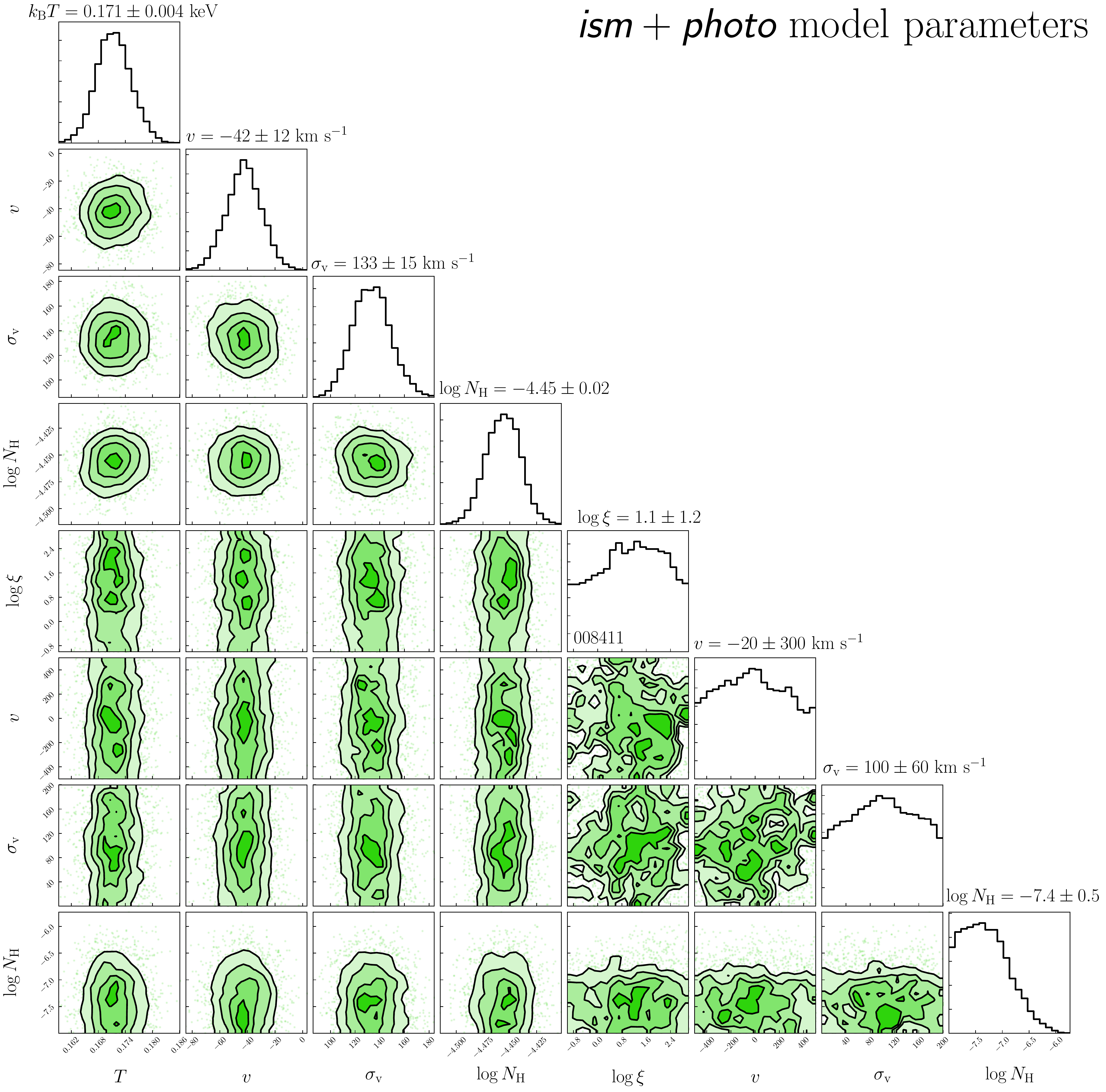}
      \caption{The posterior distribution from MultiNest is plotted with two-dimensional histograms comparing each pair of free parameters of the \texttt{hot}-model (\NH, $k_{\rm B}T$, $v$ and $\sigma_{\rm v}$), in the four upper distributions, together with the free parameters of the \texttt{xabs}-model (\NH, $v$, $\sigma_{v}$ and $\log \xi$. For clarity, we show only the $\log \xi$ of observation 008411) in the four lower distributions. The contour indicate the $1 \sigma$, $2 \sigma$, $3 \sigma$ and $4 \sigma$ confidence intervals in two dimensional space. The hydrogen column densities \NH are expressed in unit of $10^{24} cm^{-2}$.}
         \label{fig:hot_xabs_dist}
   \end{figure*}

In Figures \ref{fig:hot_dist} and \ref{fig:xabs_dist}, we display, respectively, the normalised probability distributions of the \textit{ism} and \textit{photo} model parameters computed through the Bayesian parameter inference. The estimate values with their errors are reported on the top of each panel. We also illustrate the two-dimensional distribution of the probability pairing the free parameters with each other. In both fits, we do not observe any strong covariance between the free parameters of the fit. Only the ionisation parameters ($\xi$) and the hydrogen column density of the absorber show a weak correlation visible in the $\log \xi - \log N_{\rm H}$ plots of Figure \ref{fig:xabs_dist}. \\
We also test the coexistence of collisionally ionised and photoionised gases along the line of sight fitting together the two models (\emph{ism+photo} model). Since the Bayesian approach explores the full parameter space, it can return more solid results than the $C$-stat. We show the probability distribution of the parameters for these models in Figure \ref{fig:hot_xabs_dist}. As an example, we show the ionisation parameter of observation 008411. While the parameters of the \texttt{hot} component (the four upper distributions) are well defined, the \texttt{xabs} quantities (the four lower distributions) are not constrained. Only the relative column density shows a peaked probability distribution indicating that the fraction of the possible photoionisation gas in modelling the high-ionisation lines is very small (less than 0.002). \\
Moreover, we investigate if multiple temperature interstellar hot gas (\emph{ism+ism} model) or multiple photoionised gases (\emph{photo+photo} model) can better describe the absorption features. In both cases we find that a single gaseous component dominates the fit and the contribution of the other can be considered negligible. \\
To understand which candidate among the \emph{ism}, \emph{photo}, \emph{ism+photo}, \emph{ism+ism}, \emph{photo+photo} model better fits the high-ionisation lines of \foru, we compute the Bayesian model analysis presented in the following subsection.

\subsubsection{Model comparison}
Bayesian model comparison is done by comparing models on the basis of the posterior probability of the model given the data, known as \textit{Bayesian evidence}. Using the Bayes’s rule, the Bayesian evidence ($Z$) is proportional to the prior probability for the model, $p(M)$, multiplied by the likelihood of the data ($y$) given the model, $p(y|M)$. The choice between two models, $M_1$ and $M_2$, can be made on the basis of the ratio of their Bayesian evidences. This ratio is known as Bayes factor \citep[$B_{1,2}$,][]{Wasserman00,Liddle07,Trotta08,Knuth14}. Since we choose uniform priors for different models in our analysis, the Bayes factor is defined as:
\[
B_{1,2} = \frac{Z_1}{Z_2} = \frac{p(y|M_1)}{p(y|M_2)}  \ .
\]
A large value of this ratio gives support for $M_1$ over $M_2$. Specifically, we adopt the scale of \cite{Jeffreys61} and we rule out models which show $B_{1,2}>30$ ($\log B_{1,2}>1.5$). A Bayes factor above 30 represents, indeed, a "very strong evidence" against $M_2$. The strength of this method is that it does not require models to be nested (i.e. $M_{2}$ as a special case of $M_{1}$) nor does it make assumptions about the parameter space or the data. Moreover, it automatically introduces a penalty for including too much model complexity, guarding against overfitting the data \citep{Kass95}. \\
The Bayesian statistics includes an approximation to the Bayesian evidence known as the Bayesian Information Criterion \citep[BIC,][]{Schwarz78} which is defined as ${\rm BIC} = -2\ln \mathcal{L} + k\ln N$, where $\mathcal{L}$ is the maximum likelihood, $k$ the number of parameters of the model and $N$ the number of data points used in the fit.\\
An alternative model comparison are the information-theoretic methods, pioneered by \cite{Akaike74} with his Akaike Information Criterion (AIC), which is defined as ${\rm AIC} = -2\ln \mathcal{L} + 2k$. Similarly to the Bayesian model selection, AIC and BIC can be used to decide which model is more probable to have produced the data: the most probable model corresponds to the fit with the smallest AIC and BIC value, respectively. Both AIC and BIC include an over-fitting term by which the more complex model is disfavoured by the additional number of parameters. A description geared to astronomers can be found in \cite{Takeuchi00} and \cite{Liddle04}, while the full statistical procedure can be found in \cite{Burnham02}. \\
We compare all the models presented in Section \ref{sec:parameter_inf} ($ism$, $photo$, $ism+photo$, $ism+ism$, and $photo+photo$) using the Bayesian evidence. We list the results of the model comparison in Table \ref{tab:comparison}, where we also report the relative AIC and BIC values. An interstellar hot gas with a single temperature is the favourite scenario for describing the high-ionisation lines observed in the spectra of \foru. The result of the Bayesian analysis is, therefore, in agreement with the outcome of the $C$-statistic (see Section \ref{sec:cstat}).

%
\begin{table}
\caption{Model selection results for 4U 1820-30.}
\label{tab:comparison}     
\centering          
\begin{tabular}{ c c c c  }     
\hline\hline       
\noalign{\vskip 0.75mm}
Model & $\log B$ & $\Delta \rm{AIC}$ & $\Delta \rm{BIC}$  \\ 
\noalign{\vskip 0.75mm}
\hline                  
$ism$               & $0.0$   & $0$     & $0$   \\
$ism+photo$         & $+1.3$  & $+19$   & $+40$  \\
$ism+ism$           & $+25.5$ & $+156$  & $+204$ \\
$photo$             & $+77.3$ & $+351$  & $+363$ \\
$photo+photo$       & $+90.1$ & $+389$  & $+421$ \\

\noalign{\vskip 0.75mm}
\hline
\end{tabular}
\tablefoot{The last three columns show the model comparison based on log-evidence, AIC and BIC. The log-evidence is normalised to the maximum value found, whereas both AIC and BIC are normalised to the minimum value which indicates the preferred model. Models with $\log B > 1.5$ or $\Delta AIC (\Delta BIC) >10$ can be ruled out as a plausible model that generates the data \citep{Jeffreys61,Burnham02}.}
\end{table}

%
\begin{table*}[t]
\caption{Parameter values estimated through the Bayesian analysis.}
\label{tab:xabs_hot_bayesian}
\tiny     
\centering          
\begin{tabular}{ c c c c c @{\hspace{0mm}}c@{\hspace{5mm}} c c c c c c }     
\hline\hline       
\noalign{\vskip 0.75mm}
\multirow{4}{*}{model} & \multicolumn{4}{c}{\texttt{hot}} & & \multicolumn{6}{c}{\texttt{xabs}}  \\ \cline{2-5}\cline{7-12}
\noalign{\vskip 0.5mm}
       & \NH & $k_{\rm B}T$ & $\sigma_{\rm v}$ & $v$   &    & \NH & $\log \xi$ & $\log \xi$ & $\log \xi$  &$\sigma_{\rm v}$ & $v$  \\
       & $10^{19}\rm cm^{-2}$ & keV & km/s & km/s   &   & $10^{19}\rm cm^{-2}$ & $008441$ & $98$ & $2001$ & km/s  & km/s    \\
       & & & & & & & $0055134$ & $12444$ & $2006$ &  & \\
\noalign{\vskip 0.75mm}
\hline
\noalign{\vskip 0.75mm}
$ism$           & $3.9\pm0.3$ & $0.171\pm0.004$ & $132\pm16$ &  $-41\pm14$ &  & -- & -- & -- & -- & -- & -- \\
\noalign{\vskip 0.95mm}

\multirow{2}{*}{$photo$}         & \multirow{2}{*}{--} & \multirow{2}{*}{--}  & \multirow{2}{*}{--}  & \multirow{2}{*}{--} &  & \multirow{2}{*}{ $4.1\pm0.3$ }& $1.79\pm0.04$ & $1.69\pm0.06$ & $1.76\pm0.04$ & \multirow{2}{*}{$90\pm26$} &  \multirow{2}{*}{$-34\pm18$} \\
& & & & & & & $1.79\pm0.07$ & $1.70\pm0.07$ & $1.76\pm0.02$ & & \\
\noalign{\vskip 0.95mm}

\multirow{2}{*}{$ism+photo$} & \multirow{2}{*}{$3.5\pm0.3$} & \multirow{2}{*}{$0.171\pm0.004$} & \multirow{2}{*}{$133\pm15$} &  \multirow{2}{*}{$-42\pm12$} &  & \multirow{2}{*}{$4^{+6}_{-3}\times10^{-3}$} & $1.1\pm1.2$ &  $1.1\pm1.3$ &  $1.0\pm1.3$ & \multirow{2}{*}{$100\pm60$} & \multirow{2}{*}{$-20\pm320$} \\
& & & & & & &  $0.9\pm1.3$ &  $1.1\pm1.2$ &  $1.1\pm1.2$ & & \\
\noalign{\vskip 0.95mm}

\multirow{2}{*}{$ism+ism$}      & $<3.5$  & $0.3^{+6.8}_{-0.16}$   & $150^{+530}_{-29}$ & $-41^{+228}_{-125}$ &  & \multirow{2}{*}{--} & \multirow{2}{*}{--} & \multirow{2}{*}{--} &\multirow{2}{*}{--} &\multirow{2}{*}{--} &\multirow{2}{*}{--}  \\
\noalign{\vskip 0.25mm}
                                & $<3.5$  & $0.18^{+6.37}_{-0.01}$ & $151^{+520}_{-29}$ & $-41^{+225}_{-123}$ &  &  & &  &  \\
\noalign{\vskip 0.95mm}
                               
\multirow{4}{*}{$photo+photo$}  & \multirow{4}{*}{--} & \multirow{4}{*}{--}  & \multirow{4}{*}{--}  & \multirow{4}{*}{--}

                                & & \multirow{2}{*}{$<3.8$} & $1.8\pm0.9$ & $1.7\pm0.3$ & $1.8\pm0.5$ &\multirow{2}{*}{$140^{+550}_{-50}$} & \multirow{2}{*}{$-30^{+300}_{-150}$} \\
                                & & & & & & & $1.8\pm1.2$ & $1.7\pm1.0$ & $1.7\pm0.6$ & \\                         
\noalign{\vskip 0.25mm}
                                & & & & & & \multirow{2}{*}{$<4.0$} & $1.8\pm0.8$ & $1.7\pm0.6$ & $1.8\pm0.5$ &                        \multirow{2}{*}{$140^{+510}_{-40}$} & \multirow{2}{*}{$-40^{+170}_{-160}$} \\            
                                & & & & & & & $1.8\pm1.0$ & $1.6\pm0.9$ & $1.8\pm0.8$ & \\ 
\noalign{\vskip 0.75mm}
\hline
\end{tabular}
\end{table*}

\section{Discussion}
\label{sec:discussion}

\begin{figure*}
   \centering
     \includegraphics[width=\textwidth]{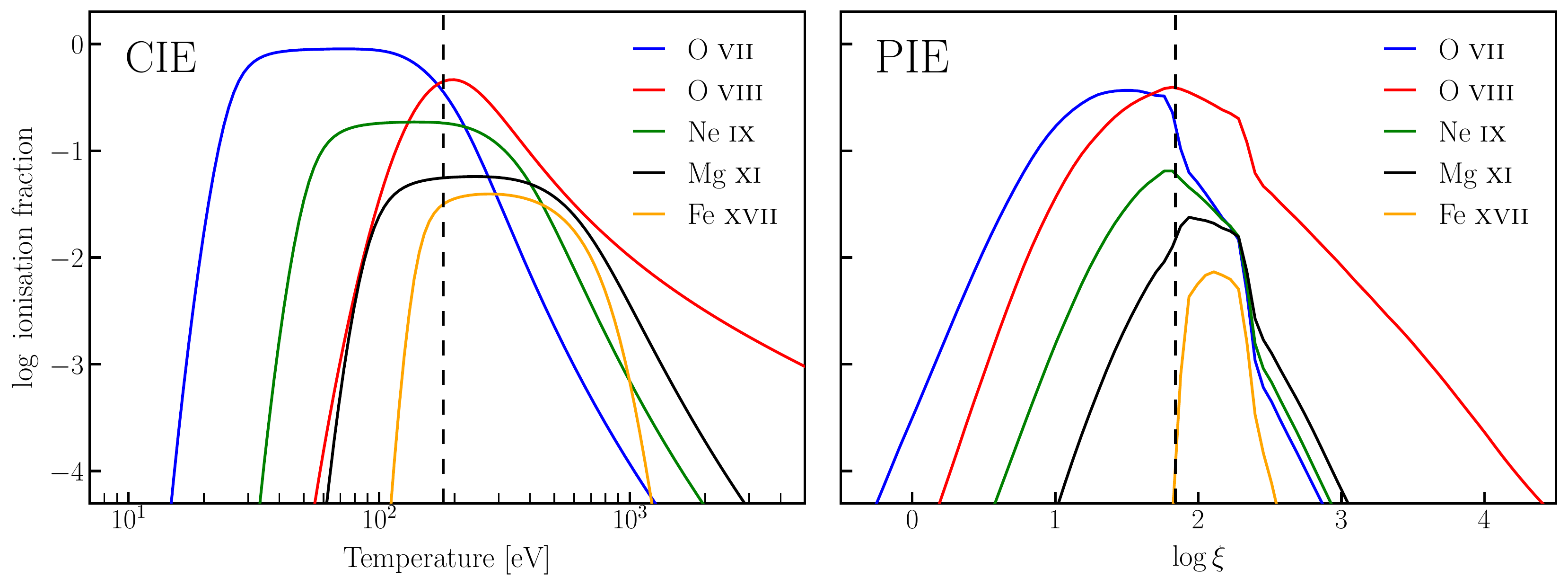}
      \caption{Ionisation fractions of oxygen, neon, magnesium and iron ions as a function of the electron temperature for collisional ionisation equilibrium plasma (CIE, $ism$-model on the left panel) and as a function of the ionisation parameter for a gas in photoionisation equilibrium (PIE, $photo$-model, on the right panel). The vertical dashed lines indicate the best values for $k_{\rm B}T$ and $\xi$ (observation 008411) obtained through the Bayesian analysis.}
         \label{fig:both_col}
   \end{figure*}

In order to determine the origin of the high-ionisation absorption lines observed in the spectra of \foru, we fit them adopting both photo-ionisation and collisional-ionisation models. In Figure \ref{fig:all_lines}, we compare the fits of the two models obtained through the Bayesian parameter inference. The \emph{ism} model reproduces all the lines better than the \emph{photo} model, in particular the \fexvii which is detected with a significance of $6.5\sigma$. This line has been previously detected by \cite{Yao06a} and \cite{Luo14}. Whereas the former propose an interstellar origin for this line, the latter conclude that the majority of the \fexvii absorbing gas arises from an intrinsic gas to the source. In our analysis, we noticed that the $photo$-model does not describe the \fexvii line (see Figure \ref{fig:all_lines}), since the fraction of the ion according to the photo-ionised model is very low.\\  
In Figure \ref{fig:both_col}, we plot the ionisation fractions of the ionised atoms for which we detected the lines in the spectrum. In particular, on the left panel, we show the relative column densities of the \emph{ism} model whereas on the right, the respective column densities for the \emph{photo} model, calculated for observation 008411. With the dashed vertical lines we indicate the best value of $k_{\rm B}T$ and $\xi$ obtained with the Bayesian analysis. The different distributions of the column densities of the ions are crucial to select which model fit the best the spectral features. For a photoionised gas with a mild photoionisation parameter ($\log \xi \sim 1.8$), iron is divided in multiple ionisation states. Consequently, the ionisation fraction of each individual Fe ion becomes low. Instead, for a collisionally ionised gas the column densities are distributed differently: the ionisation fraction is more peaked at the preferred ionisation state, which is determined by the temperature of the gas. Therefore, the detection of the \fexvii line suggests an interstellar nature for the high-ionisation lines. \\
The photoionisation nature is however statistically ruled out (see Table \ref{tab:comparison}). All the three model comparison criteria considered indicate the $ism$ as preferred model. The evidence against the $ism+photo$ model is particularly strong with the $BIC$ criterion which severely penalises complex models. In Table \ref{tab:xabs_hot_bayesian} we list the value of the parameters obtained with Bayesian analysis. In the following, we discuss in detail the results obtained for the \emph{ism} and \emph{photo} models.\\

\subsection{\it Photoionisation origin}

Both Bayesian and $C$-statistic approaches show a less statistical significance for the photoionisation modelling with respect to the collision-ionisation one. The absorber is found to have a small outflow velocity which does not support the presence of a disc wind. This velocity has to be corrected for the solar-system barycentric radial velocity of the satellite. During all the  HETG observations (obsid 1021, 1022, 6633, and 6634, 7032) this velocity was $v_{\rm{bar}}\sim30$ km/s, which explains the low outflow velocity observed. Due to its high energy resolution, the HETG instrument puts more stringent constraints on a possible flow velocity.\\

Moreover, the presence of a local ionised gas, such as a disc atmosphere, would be difficult to justify given the physical parameters of the source. Through the ionisation parameter, we can calculate the density, $n_{\rm{H}}$, if one knows where the plasma is located with respect to the X-ray source. Taking an ionising luminosity of $L_{\rm ion} = 8 \times 10^{37}\ \rm erg\ s^{-1}$ \citepalias[computed using the SED of][for the range 1-1000 Ryd]{Costantini12}, a distance $r<1.3 \times 10^{10}$ \citep[which correspondes to the size of the system,][]{Stella87} and a ionisation parameters $\xi = 62\ \rm erg\ s^{-1} cm$ (Section \ref{sec:bayesian}), the density of the plasma must be $n>7\times 10^{15}\ \rm{cm}^{-3}$, some orders of magnitude larger than the density of the typical disc atmosphere of a low-mass X-ray binary system \citep[e.g.,][]{vanPeet09}. This large density value would also imply that the filling factor, $f=N_{\rm H}/nr$, is extremely low $f < 4\times10^{-7}$, with $N_{\rm H} = 3.9 \times 10^{19}\ \rm{cm}^{-2}$. \cite{Cackett08} argued that a proper physical solution for such small filling factor is the presence of dense structures above the disc, like high-density blobs, produced by thermal instabilities. 
For example, dense clumps, possibly part of the accretion bulge observed in dipping sources, have been observed \citep[e.g.,][]{Psaradaki18}. However, their density is $n_{H} \sim 10^{13}\ \rm{cm}^{-3}$, two orders of magnitude lower than the density expected for a possible locally photoionised material. \\
Another argument against the photoionisation origin is the lack of variability of the lines observed among the different epochs \citep[Figure \ref{fig:variance} and][]{Cackett08}.\\

However, the fact that the collisional-ionisation model best represents the ionised lines of the spectra may not exclude the existence of a second component intrinsic to the source. Thus, to test the possible presence of this photo-ionised gas beside the hot interstellar plasma we compute a fit using both models. Our analysis, in particular the $C$-statistic, shows a lack of a significant contribution by photo-ionised gas towards the source for the epochs covered by the observations. Furthermore the Bayesian parameter inference shows a negligible column density of the photo-ionised gas (see Figure \ref{fig:hot_xabs_dist}).\\

A question that arises is why we do not detect any significant absorption lines from a photoionised wind/atmosphere intrinsic to \foru. Considering the SED of the ionisation source and the small size of the system, the disc/atmosphere could be fully ionised and it will be impossible to detect any absorption lines \citep{Stella87,Futamoto04}. Yet the inclination angle of the system can be a crucial factor for the non-detection of photoionised lines. \cite{DiazTrigo16} showed that the preferential detection of absorbing photoionised plasma is for system with a relative high ($i > 50\degree$) inclination. The lower inclination of the \foru \citep[$i=43\degree^{+8}_{-9}$,][]{Anderson97} system may imprint absorption lines too weak to be detected as our line of sight would go through a thinner layer of gas. \\

\subsection{\it Interstellar origin}
A single temperature hot gas can describe the absorption lines well. In specific, we observe a gas with a temperature of $\sim 1.98\times10^6$ K, which is consistent with the values obtained by previous works \citep[e.g.,][]{Yao06a} and with the temperature observed for several Galactic low-mass X-ray binaries \citep{Yao05}. Furthermore, we tested the presence of multiple temperature gases adding a supplementary \texttt{hot} component. One may expect variation in physical, and possibly chemical, properties of the hot gas along the line of sight to \foru which crosses several bubbles such as the Local Bubble around the Sun \citep[with $T\sim10^6$ K,][]{Kuntz00} and Loop I Bubble \citep[with $T\sim3.5\times10^{6}$ K,][]{Miller08}. 
However, from our modelling we do not find multiple gases with different temperatures. The opacity of the gas contained in local bubbles is too low to be detected in the spectrum. Therefore, for the line of sight towards \foru, we assume a single hot plasma component which is consistent with the scenario suggested by \cite{Hagihara11}. In addition, this uniform hot coronal gas is also in agreement with the presence of a hot and thick interstellar disc, as suggested by extragalactic source observations \citep{Yao09,Hagihara10}. \\ 
Through the Bayesian data analysis we are able to evaluate the velocity dispersion of the hot coronal gas, $\sigma_{\rm v}= 132\pm16\ \kms$. Constraining this quantity is not trivial and a proper collisional ionisation equilibrium model is necessary. \cite{Yao06b} found a dispersion velocity\footnote{\cite{Yao06b} defined the dispersion velocity ($b_{v}$) as $b_{v} = \sqrt{2}\cdot\sigma_{v}$)} between 117 \kms and 262 \kms for the hot gas along the line of sight towards \foru. Our velocity dispersion estimate is also in agreement with the dispersion velocities observed from the survey of the \ovi ($\sigma_{\rm v} = 50-200\ \kms$) with the Far UltraViolet Spectroscopic Explorer, \fuse \citep{Otte06}. The \ovi line is often used as tracer of gas with a peak temperature of $T\approx3\times10^5$ K. Such intermediate-temperature gas is expected primarily at the interface between the cool/warm clouds and the hot coronal gas \citep{Sembach03}. Finally, similar values of velocity dispersion and flow velocity have been also found by \cite{Luo18}, where they analysed the \ovii absorption line. \\
For a collisionally ionised gas with a temperature of $T\sim 1.98\times10^6$ K, the expected thermal broadening is $\sigma_{\rm v} = 0.0321\sqrt{T} = 43.5 \ \kms$. This value is too low to explain the observed velocity dispersion. This velocity cannot be accounted for by differential Galactic rotation either. The lines are, indeed, far wider than expected if the absorption came from a smoothly distributed ISM corotating with the disc of the Milky Way \citep{Bowen08}. The observed velocity dispersion supports, instead, the picture of a turbulent hot coronal gas mixed up by shock-heated gas from multiple supernova explosions. The numerical simulations of \cite{deAvillez05} show hot gas arises in bubbles around supernovae, which is then sheared through turbulent diffusion, destroying the bubbles and stretching the hot absorbing gas into filaments and vortices that dissipate with time. Furthermore the hot gas is involved in systematic vertical motion as it streams to the halo at a speed of $100-200$ \kms. This kinetic energy can be transformed into disordered, turbulent motions, resulting in higher turbulent velocities near the halo \citep{Kalberla98}. \\
Comparing the hydrogen column densities of the different components detected along the line sight of the source, we observe the following gas mass fraction: cold $\sim 89\%$, warm $\sim 8\%$ and hot $\sim 3\%$. These values are very similar to the mass fractions found by \cite{Gatuzz18} for 4U 1820-30 ($85\%$, $10\%$ and $5\%$) and they agree with findings for different lines of sight of Galactic sources. For example, \cite{Pinto13}, analysing the high-resolution X-ray spectra of a sample of Galactic X-ray binaries, found that, in average, the $\sim 90\%$, $\sim 7$ and $\sim 3\%$ of the ISM mass correspond to a cold, warm, hot plasma, respectively.

\subsection{Abundances of the hot coronal gas}
We further investigate the abundances of oxygen, neon, magnesium and iron for the hot phase of the interstellar medium. For these elements, we detect high-ionisation lines. Using the Bayesian parameter inference we obtain abundances close to their protosolar values: $A_{\rm O} = 0.90 \pm 0.06\ A_{\odot}$, $A_{\rm Ne} = 1.16 \pm 0.08\ A_{\odot}$, $A_{\rm Mg} = 1.9 \pm 0.4\ A_{\odot}$, and $A_{\rm Fe} = 1.07 \pm 0.11\ A_{\odot}$, keeping as references the abundances tabulated in \cite{Lodders10}. \\ 
The hot phase abundances of oxygen and neon are consistent with the characteristic depletion of these elements in the interstellar medium. Since neon is a noble element, it is very unlikely to be depleted in dust grains in all the medium phases and oxygen shows low depletion values in the interstellar medium \citep{Jenkins09}. Iron and magnesium are, instead, highly depleted from the gas phase in the interstellar medium \citep[e.g.][]{Jenkins09,Rogantini18,Rogantini19,Rogantini20}. The depletion of iron, in particular, remains high even in harsh environments \citep{Whittet02}. However, the observed abundances for magnesium and iron suggests that in the hot coronal gas along the line of sight towards \foru, these two elements are mostly present in the gas phase. Thus, this support the scenario where all the interstellar dust in this very hot phase ($T\gtrsim10^{5.5}$ K) is destroyed by frequent shocks during the dust grain processing in the interstellar medium. This interpretation may not be entirely unique. Reports of an overabundance of heavier elements relative to oxygen in neutral matter toward the Galactic centre would elevate the contribution of these elements in the hot phase as well. A comparison with different lines of sight especially towards sources at high latitudes, is therefore essential before drawing conclusion.

\section{Summary}
\label{sec:summary}
Motivated by defining the origin of the high ionisation absorption features present in the spectrum of the ultracompact system \foru, we systematically analyse all the observations present in the \chandra and \xmm archives: 5 \chandra/HETG spectra together with 2 \chandra/LETG and 2 \xmm/RGS observations. We study the soft X-ray energy band covering the main high-ionisation absorption lines: namely, \mgxi, \neix, \fexvii, \ovii He$\alpha$, \ovii He$\beta$, \oviii Ly$\alpha$ and \oviii Ly$\beta$. We adopt realistic plasma models to fit simultaneously the multiple lines: in particular we use the \spex model \texttt{xabs} to reproduce a photo-ionised gas and \texttt{hot} for a thermal gas in collisional ionised equilibrium state. A Bayesian framework is used to model the spectral absorption features. Bayesian data analysis provides a robust approach to infer the model parameters and their uncertainties and offers a solid model comparison.\\
Both the $C$-statistic and Bayesian data analysis show the presence of hot coronal gas with a temperature of $T\sim1.98 \times 10^{6}\ \textrm{K}\ (k_{B}T=0.171\pm0.004\ \rm{keV})$ along the line of sight towards \foru. This hot interstellar gas is responsible for the high-ionisation absorption lines detected in the spectra. We summarise our main results as follows:
\begin{itemize}
\item The mean centroids of the absorption lines are consistent with their rest frame wavelengths. The small line of sight velocity observed is within the uncertainty due to the dynamic radial velocity of the observer. As previous works already concluded \cite[e.g.,][]{Futamoto04}, an outflowing disc wind can be ruled out as possible ionised absorber. 
\item The lack of variability of the lines through the multiple observations shows also that the absorber is independent from the activity of the source suggesting a non-local origin for the gas.
\item In the spectra of \foru we detect the \fexvii line at $15.012$ {\AA} with a significance of $6.5 \sigma$. This line is best reproduced by a gas in collisional ionisation equilibrium. In contrast, for an absorber photoionised by the source the contribution of \fexvii is not high enough to justify the strength of absorption line detected.
\item We constrain the turbulent velocity of the hot gas to be $\sigma_{\rm v}= 132\pm16\ \kms$ which is likely driven by supernova shock-waves.
\end{itemize}

\begin{acknowledgements}
We would like to thank the anonymous referee for the useful suggestions. DR, EC, and MM are supported by the Netherlands Organisation for Scientific Research (NWO) through \emph{The Innovational Research Incentives Scheme Vidi} grant 639.042.525. The Space Research Organization of the Netherlands is supported financially by NWO. This research has made use of data obtained from the Chandra Transmission Grating Catalog and archive (\url{http://tgcat.mit.edu}), and software provided by the \emph{Chandra} X-ray Center in the application package CIAO. We also made used of the \xmm Scientific Analysis System developed by a team of scientist located at ESA's \xmm Science Operations Centre and at the \xmm Survey Science Centre. We are grateful to M. Díaz Trigo for a useful discussion on the photoionised atmosphere of \foru. We thank A. Dekker and D. Lena for reading an early draft of the manuscript and for providing valuable comments and suggestions. We also thank I. Psaradaki for her input on the \xmm data reduction and J. de Plaa for his support on the development of the program.
\end{acknowledgements}

%
%

\bibliographystyle{aa} 
\bibliography{../bibliography/biblio}

\begin{appendix}

\end{appendix}

\end{document}